\documentclass[
aps,%
12pt,%
final,%
notitlepage,%
oneside,%
onecolumn,%
nobibnotes,%
nofootinbib,%
superscriptaddress,%
noshowpacs,%
centertags]%
{revtex4}%
\usepackage{geometry}
\usepackage{graphics}

\begin{document}
\selectlanguage{english}

\title{A role of parton charges and masses
in the exclusive photon-photon production of meson pairs}

\author{\firstname{A.~V.}~\surname{Berezhnoy}}
\email{aber@ttk.ru}
\affiliation{SINP MSU, Moscow, Russia}%
\author{\firstname{A.~K.}~\surname{Likhoded}}
\email{likhoded@mx.ihep.su}
\affiliation{IHEP, Protvino, Russia}

\begin{abstract}
The exclusive photon-photon production of the $K$-meson pairs 
has been analyzed within the partonic model of QCD for the kinematical 
conditions of BELLE experiment. 
The  cross section dependences on partonic masses, charges, and
a shape of $K$-meson wave function have been studied for the process
under discussion.  

\end{abstract}

\maketitle

\section{Introduction}
Not long ago the angular distributions for 
the process   $\gamma\gamma\to K^+K^-$ 
have been measured by BELLE collaboration at energies 
 2.4-4.1~GeV \cite{Belle}. The obtained results are in 
 satisfactory agreement with the predictions of article \cite{Brodsky}, 
 where the  amplitude of  exclusive production of meson pairs 
had been performed as the amplitude $T_H$ of hard production of
four partons convoluted with the parton amplitude distributions 
$\phi(x,Q)$ of the final mesons ($x$ is a momentum fraction 
carried by valence parton and $Q$ is a interaction scale): 
\begin{equation}
A = \int\!\!\!\int dx\, dy\,\phi^*(x,Q) \phi^*(y,Q) T_H.
\label{fact}
\end{equation} 
The amplitude $T_H$ of a hard subprocess of 
four massless quark production
 $\gamma \gamma \to q \bar q q \bar q$ was calculated in the frame work
of perturbative QCD. It was rather  difficult task because one need to take 
into account 24 Feynman diagrams of $O(\alpha^2\alpha^2_s)$.

The following parameterizations were used for $\phi(x,Q)$:
\begin{equation} 
\phi(x,Q)\sim x(1-x),
\label{common_distr}
\end{equation}   
\begin{equation} 
\phi(x,Q)\sim [x(1-x)]^{\frac{1}{4}},
\label{broad_distr}
\end{equation}   
\begin{equation} 
\phi(x,Q)\sim \delta\left(x-\frac{1}{2}\right).
\label{delta_distr}
\end{equation}   
Transverse momenta of parton inside mesons, as well as partonic masses were
neglected. 

Now-days the wave function parametrization  (\ref{common_distr}) is 
the most used for the study of processes with light mesons within
the light-cone thechnique.
The parametrization (\ref{broad_distr}) was used 
in work \cite{Brodsky}, as an example of a function with 
a large partonic momentum dispersion.  On the contrary,  
the wave function (\ref{delta_distr})
describes a meson with fixed momenta of constituent partons.
The latter parametrization has been used many times to study the processes with
heavy quarkonia (see for example a set of papers devoted to   
 $B_c$-production in different interactions \cite{Bc}). Also the
parameterization  (\ref{broad_distr}) has been used by us to estimate 
the production cross sections
and the charge asymmetries in the charm meson photoproduction \cite{BKL}. 

One of the aims of this study is to estimate the role 
of partonic masses in the inclusive photonic production of meson pairs. 
For this purpose we have taken into account the
mass corrections in the hard process amplitude  $T_H$. 

The other aim of the research is to note one more time, that the values of
 partonic charges  influence strongly the shapes of differential distributions.
Our calculations, as well as the results of \cite{Brodsky} are indicative of 
the cardinal difference between the angle distributions for charged meson
pairs and ones for neutral meson pairs.  This difference is clearly seen 
from the formula for the cross section distribution over cosine of 
azimuthal angle $\Theta$ in c.m.s. for the production of pseudoscalar meson
pairs, which had been  obtained in \cite{Brodsky}:
\begin{equation}
\frac{d\sigma}{d\cos\Theta}\sim \frac{(e_1-e_2)^4}{\sin\Theta^4}+
\frac{2e_1e_2(e_1-e_2)^2}{\sin\Theta^2}g[\Theta, \phi]+ 
2(2e_1e_2)^2g^2[\Theta, \phi],
\label{brodsky}
\end{equation}
where $g[\Theta, \phi]$ is a function of angle and of $\phi(x)$,
$e_1$ and $e_2$ are the quark charges (the charges of produced
mesons are equal to  $\pm (e_1-e_2)$). 
It is  worth to underline that the both terms in (\ref{brodsky}) can
contribute essentially to  $K^+ K^-$-pair production process. That is why  
the  approximation to the cross section in the form of $1/\sin\Theta^4$,
which was used in \cite{Belle},  does not take into account all features
of the model under discussion.
For example, for  the $\delta$-shaped amplitude $\phi(x)$ and for
massless partons  $g[\Theta, \phi]\equiv 1$ and the cross section of   
 $K^+K^-$-pair depends on  $\Theta$ as follows:
\begin{equation}
\frac{d\sigma}{d\cos\Theta}\sim \frac{1}{\sin\Theta^4}-
\frac{4}{9\sin\Theta^2}.
\label{brodsky_delta}
\end{equation}
Let us to note that the term proportional to $1/\sin^2\Theta$ allows to
slightly improve  the description of experimental data published in  \cite{Belle}.      

As it was already mentioned above, the model under discussion predicts the 
essential difference between the cross section distributions for
neutral meson pair production and for charged meson pair production. 
It is especially interesting because the other models \cite{Vogt} 
does not predict an essential dependence of the differential cross section on
parton charges.
    
The third aim of the present work is to  find-out the influence of a shape  
of $\phi(x,Q)$ on the behavior of cross section distributions for 
massive partons, as well as massless ones.

\section{An amplitude distribution and a mass of a system constituted from
 two partons}
 
Before to state the main results of the study let us return to the discussion 
of amplitude distribution shape to make two remarks about this.  
As it is mentioned above, the approximation (\ref{common_distr}) is the
most used
for the case of light mesons constituted from massless partons.
The validity  of  parametrization (\ref {common_distr})
for calculations is widely discussed in the literature \cite {Amp_distr}, 
and we do not consider here this question. 
Our observation concerns the other aspect of the problem.  
Let us suppose that the valence quarks $q_1$ and $q_2$ have
 the effective masses
 $m_{q_1}$ and $m_{q_2}$. Let us also suppose that these partons carry 
practically the total momentum of meson.  Then neglecting their transverse
momenta we obtain a following expression 
for the mass squared of such two parton system:
 
\begin{equation} 
m^2(x)=\Biggl(\sqrt{x^2 p^2+m_{q_1}^2}+\sqrt{(1-x)^2 p^2
+m_{q_2}^2}\Biggr)^2-p^2,
\end{equation}   
where  $x$ is a fraction of meson momentum  $p$ carried away by
the parton  $q_1$.  At large momenta  $m^2(x)$ can be represented as 
\begin{equation}
m^2(x)\Bigl|_{p\to \infty}\Bigr.=\frac{m_{q_1}^2}{x}+\frac{m_{q_2}^2}{1-x}.
\label{mx2_limit}
\end{equation}

It is easy to show that the function $m(x)$ reaches the minimum value
$m_{q_1}+m_{q_2}$ at $x=m_{q_1}/(m_{q_1}+m_{q_2})$.

If $m_{q_1}=m_{q_2}$ then $m^2(x)\sim 1/(x(1-x)) $ and the amplitude
 (\ref{common_distr}) depends on $x$ as $1/m^2(x)$. Thus, one could suppose
that for unequal masses  $m_{q_1}$ and $m_{q_2}$ a following
 heuristic expression for the parton amplitude can be used: 
\begin{equation}
\phi(x)\sim \frac{1}{m^2(x)}\sim 
\Biggl(\frac{r^2}{x}+\frac{1}{1-x}\Biggr)^{-1},
\label{kaon}
\end{equation}
where $r=m_{q_1}/m_{q_2}$. At least, for 
the reasonable value of parameter $r=2$ the received distribution  is
practically indistinguishable from one represented in work  \cite{Braun} 
for  $K$-meson. 
 
The other remark  is that even for a heavy quarkonium the value of $m(x)$
can not be identify with the quarkonium mass because this mass can not
depend on $x$. Nevertheless, we think that a physical meaning have only 
the values of $x$,  at which the difference between $m(x)$ and 
the meson mass $M$ not larger than a coupling energy value  $\Lambda$.
If the both quarks are heavy then $M\approx m_{q_1}+m_{q_2}$, $M\gg\Lambda$,
and we arrive at the following  inequality: 
\begin{equation}
m(x)-m_{q_1}-m_{q_2}\lesssim \Lambda.
\label{mx_range}
\end{equation}
In addition to that, if $m_{q_1}=m_{q_2}$, then from  (\ref{mx2_limit}) and (\ref{mx_range})
we obtain:
\begin{equation}
\frac{1}{2}-\frac{1}{2}\sqrt{\frac{\Lambda}{m_q}}
\lesssim x \lesssim
\frac{1}{2}+\frac{1}{2}\sqrt{\frac{\Lambda}{m_q}}.
\label{x_range}
\end{equation}

Using (\ref{x_range}) one can easy to estimate a region 
of reasonable value of $x$  both for charmonium 
($0.3 \lesssim x \lesssim 0.7$) and bottomium 
($0.35 \lesssim x \lesssim 0.65$). 

Thus, following  \cite{Brodsky}, one can essentially simplify
the problem if  suppose that the amplitude of exclusive 
production of meson pair can be subdivided into the hard amplitude of parton 
production and the soft adronization amplitude.    
But such a factorization cause the vagueness of meson mass.
Different values of  $x$ correspond to different values of a 
mass of system with two partons. In other words, 
we replace a meson with a  fixed mass value to 
a wave package  constituted by partons. Such an object can not be
characterized by the definite mass value. Nevertheless, a rather wide 
region of $x$ variable exits where the mass of two parton reach 
physically reasonable values  as for light mesons, 
 as for heavy ones.  Therefore the factorization proposed in 
 \cite{Brodsky} is physically justified.
   
\section{Calculation results}
In the present work we do not concern the problem of absolute normalization of
the cross section and discuss only the shape of cross section 
distribution with respect to $\cos \Theta$ depending on the following factors:
\begin{enumerate}
\item a value of meson charge,
\item values of effective parton masses,
\item a shape of parton amplitude $\phi(x)$.
\end{enumerate}
  
At the present time there are many work  \cite{Amp_distr} devoted 
to study of different parameterizations of parton amplitude and their behavior. 
Nevertheless,  as it was experienced by previous calculations, 
the shape of parton amplitude can not be reconstructed in details from the 
behavior of cross section or decay width.  The process under discussion
is not an exception and the differential cross section of the meson pair
production allows us to determine only qualitative characteristics of the 
parton amplitude $\phi(x)$. That is why use the rough model represented in 
\cite{Brodsky} to obtain our predictions.

The cross section calculations for   $K$-meson production have been done
within the following assumption sets about a parton amplitude shape and
parton masses: 
\begin{enumerate}
\item  $s$-quark mass equals to $m_s=0.2$~GeV, the masses of $u$- and
 $d$-quarks equal to $m_q=0.1$~GeV  
and the momentum fractions carried  by the partons are fixed: 
$\phi(x)\sim \delta(x-m_s/(m_s+m_q))$.
\item $m_s=0.2$~GeV, $m_q=0.1$~GeV, and the parton amplitude are 
  spread as much as possible: $\phi(x)=const>0$ at $0.25<x<0.95$ and
$\phi(x)=0$ at other $x$. The region of $x$ at which the parton amplitude
reaches nonzero values has been estimated from the condition (\ref{mx_range}).
\item The parton masses are negligibly small, and  $\phi(x)\sim \delta(x-2/3)$. 
\item The parton masses are negligibly small, and $\phi(x)$ have a shape
(\ref{kaon}). 
\end{enumerate}

Of course these extreme variants  are not realized in the nature.  
However the study of the itemized assumption sets allows to better understand
a new experimental data.   

Fig.~1a presents the normalized differential cross sections of pair
production of charged $K$-mesons with respect to $|\cos\Theta|$ 
for the process $\gamma\gamma\to K^{+}K^{-}$ at $2.65$~GeV for the 
supposition sets under discussion: (1) (solid curve), 
(2) (dashed curve), (3) (dotted curve), and (4) (dashed-dotted curve). 
The experimental data of Belle Collaboration are also given in Fig.~1a for
comparison. All four sets describe data fairly well. 
Surprisingly, that the most fair description has been achieved
for the variant (3) which  can be approximated by
the formula  (\ref{brodsky_delta}) with a high accuracy. However 
one can not  come to a final decision. 
 
Fig.~1b  presents the  differential cross sections of pair
production of charged vector and pseudoscalar $K$-mesons 
($\gamma\gamma\to K^{+}K^{*-}(K^{*+}K^{-})$) normalized to
the cross section value of pair production of charged pseudoscalar mesons. 
It is seen from Fig.~1b that the cross section  of the process under 
discussion depends on $\Theta$ as $\sim \cos^2\Theta/\sin^4\Theta$.
Also one can see that the relative yield of pseudoscalar-vector pairs  for
the ``spread'' parton amplitudes is essentially smaller than for   
the $\delta$-shaped one.

As it is seen from Fig.~1c for  the case of massless partons
the yield of the vector-vector pairs is smaller than for the
case of massive ones. On comparing Fig.~1c, 1a, and 1b,  it is clear, 
that the cross section of the process  $\gamma\gamma\to K^{*+}K^{*-}$ 
depends on  $|\cos \Theta|$  weaker  than the cross sections of the processes           
 $\gamma\gamma\to K^{+}K^{-}$ and $\gamma\gamma\to K^{+}K^{*-}(K^{*+}K^{-})$.

The differential production cross sections of the pairs of neutral pseudoscalar
$K$-mesons are crucially differ from  the cross sections obtained for the
charged pair production. The  cross section for process 
$\gamma\gamma\to K^{+}K^{-}$ strongly depends on $|\cos \Theta|$. 
On the contrary, the distribution over $|\cos \Theta|$ for the process
 $\gamma\gamma\to K^{0}\bar K^{0}$ is practically flat for 
the $\delta$-shaped parton amplitude and slightly curved for the spread 
 parton amplitude (see Fig.~1d).

It is clear from Fig.~4e, that the process 
$\gamma\gamma\to K^{0} \bar K^{*0}(K^{*0} \bar K^{0})$ is 
strongly suppressed in the case of massless partons.
For the scheme (1) with massive partons and fixed parton momenta
the cross section distribution over  $|\cos \Theta|$ with fair accuracy
can be approximated by the function $|1/\sin^3\Theta|$.
A spreading of parton amplitude within scheme (2) leads to the increase of
the relative $ K^{0} \bar K^{*0}$-pair yield and to 
the slight alteration of the distribution shape.

The values of parton charges influence strongly  the behavior
of cross section distribution 
of pseudoscalar-pseudoscalar and pseudoscalar-vector pairs, 
whereas the distribution shape of vector-vector pair production practically 
does not depend on  parton charges, as  
it is clear from comparing Fig.~1f and Fig.~1c were the cross section
distributions over  $|\cos \Theta|$ are shown 
the processes  $\gamma\gamma\to K^{*0}\bar K^{*0}$ and 
$\gamma\gamma\to K^{*+}K^{*-}$, correspondingly. 

Fig.~2 presents the same distributions as Fig.~1 but at the
photon-photon interaction energy $3.15$~GeV. 
One can see that there is no essential difference between the distributions
at $2.65$~GeV (Fig.~1) and at $3.15$~GeV (Fig.~2).

\section{Conclusions}
Our calculations show that the shapes of differential cross section  of
the pseudoscalar meson pair production in the photon-photon interaction 
is determined in general by the meson charge. 
For the process  $\gamma\gamma\to K^{+}K^{-}$ the distribution over
$\cos \Theta$ have a strongly marked  peripheral character, whereas 
the analogous distribution for the process   
$\gamma\gamma\to K^{0}\bar K^{0}$ is practically flat.
It is worth to note that the partonic mass values and 
the shape of parton amplitude influence weakly the differential cross section
behaviour. 

The differential cross section shapes of pseudoscalar-vector production 
are determined by the meson charge too. But the relative yield 
of pseudoscalar-vector pairs and  vector-vector pairs depends essentially on 
the  parton amplitude parameterization, as well as on the parton mass values.
It is interesting that the prediction of cross section value for the process 
$\gamma\gamma\to K^{0} \bar K^{*0}(K^{*0} \bar K^{0})$ can be essentially 
reduced by using a spread parton amplitude, whereas  
the cross section value  for the process  
$\gamma\gamma\to K^{+} K^{*-}(K^{*+} \bar K^{-})$ is more sensitive to  the
choice of parton masses values and shape of parton amplitude.
For the massless partons the cross section of the latter process 
is negligibly small independently on the chosen parton amplitude
parametrization. 

The  vector-vector pair production have a weakly marked peripheral  character. 
The relative yield of vector-vector pair, as well as the shape of 
cross section distributions weakly depend on suppositions about both parton 
distributions and parton mass values.

The  neutral vector-vector pair production have a marked peripheral  character
for spread parton amplitudes and is practically isotropic for 
the  $\delta$-shaped amplitude.

To conclude, the present research shows  that all tree discussed factors
(the meson charge value, the shape of parton amplitude, and the values of 
parton masses) can essentially influence the exclusive meson pair production
in the photon-photon interaction.
Nevertheless, the role of meson charge value must be especially stressed,
because a change of meson charge leads to the most crucial change in the
meson production.  That is why it should  be very interesting 
to compare the experimental results on the  pair production of charged 
$K$-mesons \cite{Belle} with experimental results on the pair production
of neutral $K$-mesons.

This work was supported in part by the Russian Foundation for Basic Research 
(project no.~04-02-17530) and CRDF (grant no.~M0-011-0).
It was also funded within the Program for Support of Leading Scientific Schools
(grant no.~1303.2003.2).

\newpage

\begin{figure} 
{\vspace*{-2cm}
\centering \resizebox*{\textwidth}{!}{\includegraphics{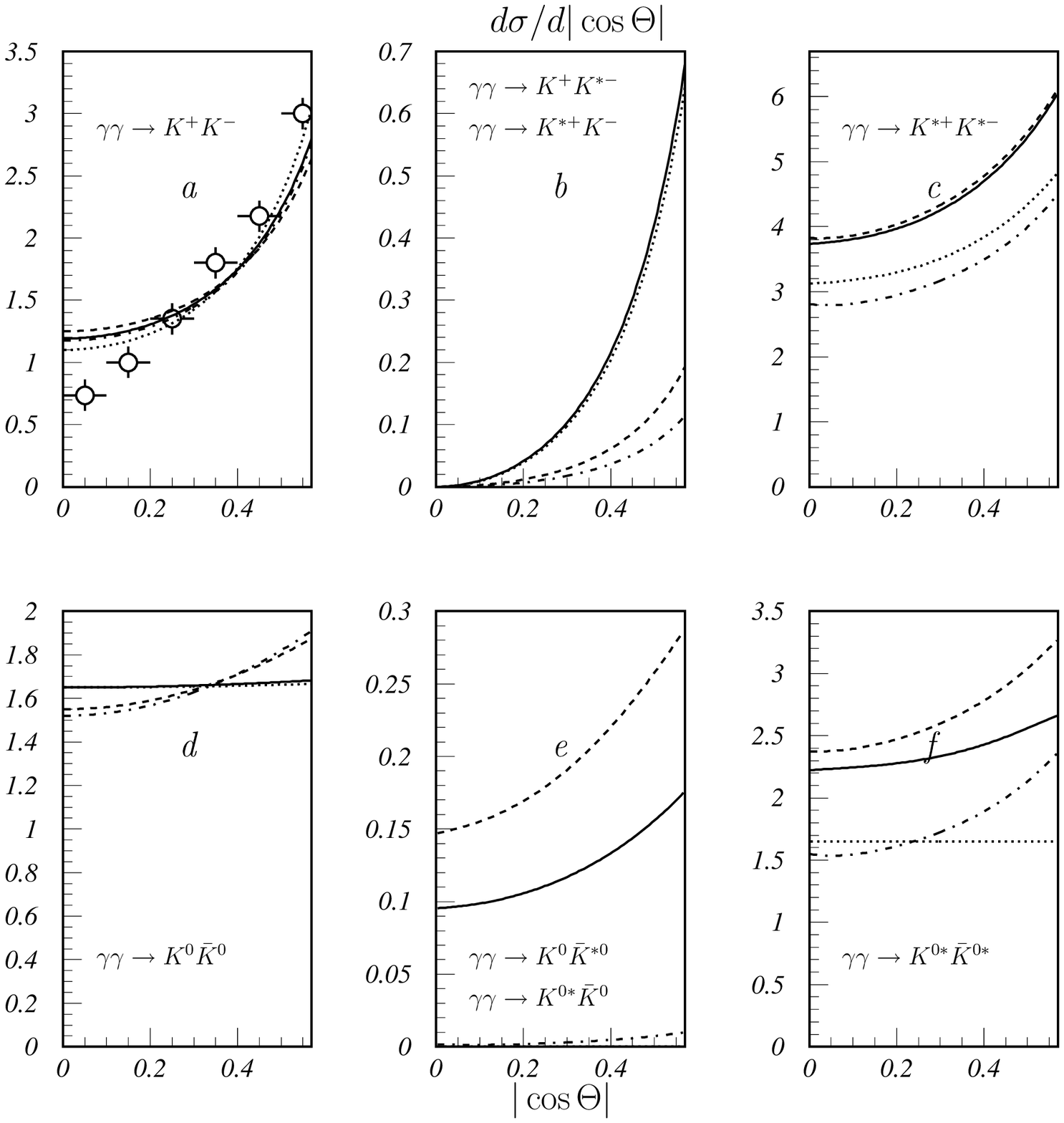}}

\vspace*{-6cm} \par}
\setcaptionmargin{5mm}
\onelinecaptionsfalse
\captionstyle{normal}
\caption{ The differential cross sections of $K$-meson pair production in
 the photon-photon interaction at $2.65$~GeV with respect to 
 \(|\cos \Theta|\) for the assumption sets under discussion:
(1) \(m_s=0.2\)~GeV, \(m_u=m_d=0.1\)~GeV, 
\(\phi(x)\sim \delta(x-m_s/(m_s+m_q))\)
(solid curve); (2) \(m_s=0.2\)~GeV, \(m_u=m_d=0.1\)~GeV, 
\(\phi(x)=const>0\) at \(0.25<x<0.95\) and \(\phi(x)=0\) at other
\(x\) (dashed curve); (3) the parton masses are negligibly small, 
\(\phi(x)\sim \delta(x-2/3)\) (dotted curve); 
(4) the parton masses are negligibly small,  \(\phi(x)\) have a shape
(\ref{kaon}) (dashed-dotted curve). See the text for the detailed explanation.
}
\label{kk2.65.ps}
\end{figure}

\begin{figure} 
{\vspace*{-2cm}
\centering \resizebox*{\textwidth}{!}{\includegraphics{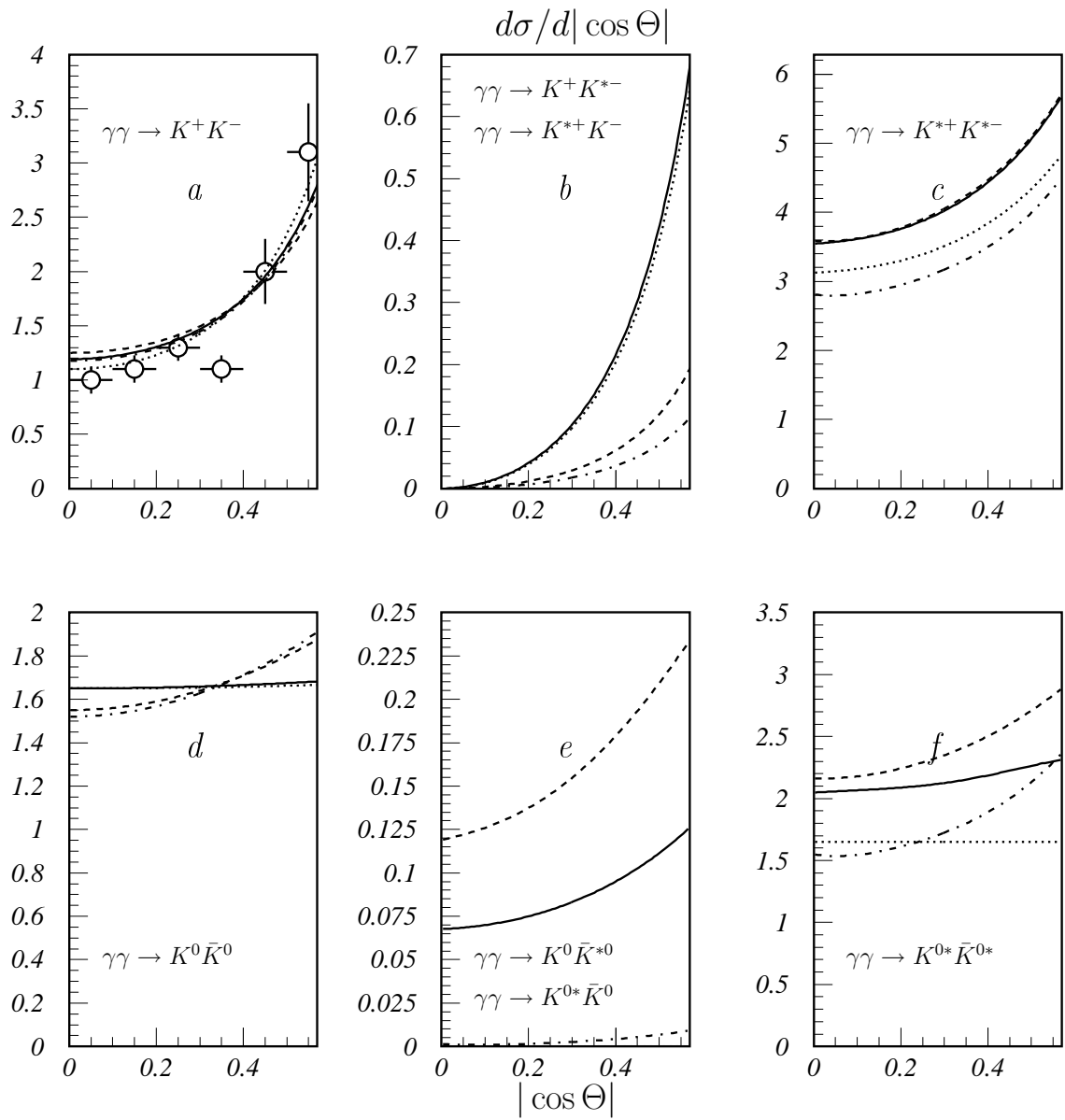}}

\vspace*{-6cm} \par}
\setcaptionmargin{5mm}
\captionstyle{normal}
\caption{ The same as in Fig.~1, but for the interaction energy $3.15$~GeV.}
\label{kk3.15.ps}
\end{figure}

\end{document}